\documentstyle[12pt]{article}
\begin{document}

\date{}

\title{{\Large {\bf CHAOS IN THE $Z(2)$ GAUGE MODEL ON A GENERALIZED BETHE
LATTICE OF PLAQUETTES}}}
\author{{\bf N.S. Ananikian}\thanks{e-mail: ananik@jerewan1.yerphi.am} ,
{\bf S.K. Dallakian},  
{\bf N.Sh. Izmailian} \thanks{e-mail:  izmailan@vxc.yerphi.am} 
\thanks{Present address: Academia Sinica, Institute of Physics,
Nankang, Taipei, Taiwan 11529 ROC } , \\
{\normalsize Department of Theoretical Physics, Yerevan Physics Institute,}\\
{\normalsize Alikhanian Br.2, 375036 Yerevan, Armenia}\\
{\bf K.A. Oganessyan}\thanks{e-mail: kogan@hermes.lnf.infn.it} 
\thanks{Permanent address: Yerevan Physics Institute, Alikhanian Br.2, 
375036 Yerevan, Armenia}\\
{\normalsize LNF-INFN, C.P. 13, Via Enrico Fermi 40, I-00044 Frascati 
(Rome), Italy}} 
\maketitle

\begin{abstract}
We investigate the $Z(2)$ gauge model on a generalized Bethe lattice with
three plaquette representation of the action. We obtain the cascade of phase
transitions according to Feigenbaum scheme leading to chaotic states for
some values of parameters of the model. The duality between this gauge model 
and three site Ising spin model on Husimi tree is shown. The Lyapunov 
exponent as a new order parameter for the characterization of the model 
in the chaotic region is considered. The line of the second 
order phase transition, which corresponds to the points of the first period 
doubling bifurcation, is also obtained.
\end{abstract}

\newpage

The thermodynamics of the phases and the order of the transitions plays 
an important role in many branches of physics. It is widthly 
recognized that a study of QCD phase structure will help to understand 
the one of the main problems in strong interactions, which is the 
color confinement (see \cite{H1} and references therein). At sufficiently 
high temperatures ($T_c \sim 200 MeV$ ($\sim 2.32 \times 10^{12} K$)) one 
expects a phase transition in gluodynamics: a color confinement phase 
(hadronic phase) changes to a deconfinement phase (plasma phase). The 
question of the order of this finite-temperature transition with physical 
values of three quark masses is still open. The issue of the transition 
from a confinement phase at low temperatures to the deconfinement phase 
at high temperatures can be related to the issue of whether the pure glue 
vacuum of QCD is $Z(N)$ invariant like the action. As it turns out, the 
transition may be explained as a spontaneous breaking of the extra $Z(N)$ 
symmetry at finite temperatures.    
  
On the other hand, it is well known that the vacuum of the non-Abelian gauge 
theories is non integrable in the classical limit and exhibits dynamical 
chaos \cite{A7, A8, AA8} (see also \cite{AA7} for details and extended 
literature). The connection between the maximum Lyapunov exponent and the 
gluon dumping rate in a hot perturbative QCD has been obtained in Ref.
\cite{AA9}. These chaotic solutions are playing an essential role in the self 
thermalization of the hot quark-gluon plasma in heavy ion 
collision\cite{AA7}. It is important to mention also the Ref.\cite{PK}, 
where it was shown that the deconfinement phase domains in $SU(2)$ lattice 
gauge theory have a non-integral dimension near the phase transition.   

In this letter we are presenting arguments which leads to chaotic states in
the $Z(N)$ gauge theory on the hierarchical lattice. The expectation value 
of the Polyakov loop operator of the $Z(N)$ theory determines the phase 
structure of the $SU(N)$ gauge theory at a finite temperature\cite{A9,AA10}. 
In particular the restoration of the $SU(2)$ gauge symmetry in standard model
and deconfining transition in QCD at a finite temperature are consequence of
phase transitions in $Z(2)$ and $Z(3)$ models respectively. Using the
dynamical systems or recursive approaches one can obtain an order parameter
for the $Z(N)$ gauge model on a hierarchical lattices. In many cases
recursive sequence converges to a fixed point and one can obtain a
qualitatively more correct phase structure than in conventional mean field
approximations \cite{GUJ}. It is necessary to mention that for various 
reasons it is interesting to generalize lattice gauge actions by 
including larger interaction loops. For instance, the enlarged gauge 
action involving new double plaquette interaction terms were proposed and 
studied in $3d$ and $4d$ \cite{N1,N2} and there were obtained qualitative 
changes in phase diagrams. The $2d$ version of one of these lattice gauge 
models with $Z(2)$ gauge symmetry formulated on the planar rectangular 
plaquettes was reduced to the usual spin-1/2 Ising model on the square 
lattice and the point of the second order phase transition was found 
\cite{N3}. Recently, the Z(3) gauge model with double plaquette 
representation on the flat (triangular and square) \cite{N4} and on a 
generalized recursive lattice \cite{N5} is considered and reduced it to 
the spin-1 Blume-Emery-Griffiths model \cite{BEG}. 
  
In this letter we construct a $Z(2)$ gauge model with three plaquette 
representation of the action and the cascade of phase transitions 
according to Feigenbaum scheme leading to chaotic states for some values 
of parameters of the model. Then the duality between this gauge model on 
generalized Bethe lattice of plaquettes and three-site Ising spin model 
on Husimi tree \cite{AM} becomes obvious.   

The generalized Bethe lattice is
constructed by successive building up of shells. As a zero shell we take the
central plaquette and all subsequent shells come out by gluing up two new
plaquettes to each link of a previous shell. As a result we get an infinite 
dimensional lattice on which three plaquettes are gluing up to each link 
(Fig.1a). On a lattice the gluon field is 
described by matrixes $U_{x,\mu}$ which are assigned to the links of the 
lattice. The $U_{x,\mu}$ are elements of the gauge group itself and 
$U_{x,\mu} = exp(iag_0A_{\mu}(x)) $, where $g_0$ is the unnormalized 
charge, $a$ - the distance between neighboring sites. In the $Z(2)$ 
gauge model field variables $U_{ij}$ defined on the links take their 
values among the group of the two roots of unity $U_{ij}\in 
\left\{ \pm 1\right\} $. Then the gauge invariant action in the presence 
of three plaquette interaction can be written in the form 
\begin{equation}
\label{r1}S=-\beta _3\sum_{3p}U_{3p}-\beta _1\sum_pU_p, 
\end{equation}
where 
$$
U_p=U_{ij}U_{jk}U_{kl}U_{li} 
$$
is the product of the gauge variables along the plaquette contour and 
$$
U_{3p}=U_{p_n}U_{p_{n+1}}^{^{\prime }}U_{p_{n+1}}^{^{\prime \prime
}}=U_{ij}U_{jk}U_{kl}U_{li}U_{im}U_{mn}U_{nj}U_{jq}U_{qr}U_{ri} 
$$
is the minimal product of the gauge variable along the tree plaquettes
(Fig. 2b). The $\beta _1$ and $\beta _3$ are the gauge coupling constants.

The partition function of this model is  
\begin{equation}
\label{r2}Z=\sum_{\{U\}}e^{-S},
\end{equation}
where the sum is over all possible configurations of the gauge field
variables $\{U\}$. The expectation value of the central single plaquette $P$
will have the following form 
\begin{equation}
\label{r3}
P\equiv <U_{p_0}>=Z^{-1}\sum_{\{U\}}U_{p_0}e^{-S}.
\end{equation}

The advantage of the generalized Bethe lattice is that for the models
formulated on it exact recursion relation can be derived. The partition
function separates into four identical branches, when we are cutting apart
the zero shell (central plaquette) of the generalized Bethe lattice. Then
the partition function for $n{\it th}$ generation ($n \to \infty$ corresponds 
to the thermodynamic limit) can be rewritten 
\begin{equation}
\label{r4}
Z_n=\sum_{\{U_{p_0}\}}e^{\beta _1U_{p_0}}g_n^4(U_{p_0}),
\end{equation}
where the sum is over all possible configurations of the field variables
defined on the links of a zero plaquette $\{U_{p_0}\}$ and 
\begin{equation}
\label{r5}
g_n(U_{pi})=\sum_{\{U_{p_{i+1}}^{^{\prime }},U_{p_{i+1}}^{^{\prime
\prime }}\}}e^{\beta _3U_{p_i}U_{p_{i+1}}^{\prime }U_{p_{i+1}}^{\prime
\prime }+\beta _1U_{p_{i+1}}^{\prime }+\beta _1U_{p_{i+1}}^{\prime \prime
}}g_{n-1}^3(U_{p_{i+1}}^{\prime })g_{n-1}^3(U_{p_{i+1}}^{^{\prime }\prime }).
\end{equation} 
From Eq. (\ref{r5}) one can obtain: 
$$
g_n(+)=16e_{}^{\beta _3+2\beta _1}g_{n-1}^3(+)g_{n-1}^3(+)+32e^{-\beta
_3}g_{n-1}^3(+)g_{n-1}^3(-)+16e^{\beta _3-2\beta
_1}g_{n-1}^3(-)g_{n-1}^3(-), 
$$
$$
g_n(-)=16e^{-\beta _3+2\beta _1}g_{n-1}^3(+)g_{n-1}^3(+)+32e^{\beta
_3}g_{n-1}^3(+)g_{n-1}^3(-)+16e^{-\beta _3-2\beta
_1}g_{n-1}^3(-)g_{n-1}^3(-). 
$$

Note that $U_{p_i}$ takes the values $\pm 1$ and coefficients before
exponents arise because of the gauge invariance.

After introducing the variable  
\begin{equation}
\label{R6}
x_n=\frac{g_n(+)}{g_n(-)}.
\end{equation}
one can obtain the following recursion relation
\begin{equation}
\label{R7}
x_n=f(x_{n-1}),\qquad f(x)=\frac{z{\mu }^2x^6+2\mu x^3+z}{{\mu }
^2x^6+2z\mu x^3+1},
\end{equation}
where $z=e^{2\beta _3},\quad \mu =e^{2\beta _1}$. 

Through this $x_n$ one can express the average value of the central 
plaquette for $n{\it th}$ generation 
\begin{equation}
\label{R8}
P_n=\frac{8\{e^{\beta _1}g_n^4(+)-e^{-\beta _1}g_n^4(-)\}}{
8\{e^{\beta _1}g_n^4(+)+e^{-\beta _1}g_n^4(-)\}}=\frac{e^{\beta _1}x_n^4-1}{
e^{\beta _1}x_n^4+1},
\end{equation}
which is the gauge invariant order parameter in a $Z(2)$ theory for a
stable fixed point in thermodynamic limit ($n \to \infty$).

Note that by using duality relation the above recursive relation can be
obtained for the magnetization $m$ of the three site interacting Ising spin 
model on Husimi tree with Hamiltonian\cite{AM}. 
\begin{equation}
\label{R1}
-\beta H=J_3\sum_{\triangle }{\sigma }_i{\sigma }_j{\sigma }
_k+h\sum_i{\sigma }_i, 
\end{equation}
where ${\sigma }_i$ takes values $\pm 1$, the first sum goes over all
triangular faces of the Husimi tree and the second over all sites.

So, we find a relation between $Z(2)$ gauge model with three plaquette 
representation of action on a generalized Bethe lattice and the three site 
interacting Ising spin model on Husimi tree. The duality brings to the 
following correspondence:

\begin{center}
$U_{p_i}\Leftrightarrow \sigma _i\qquad \qquad \qquad \qquad \qquad \qquad
\qquad \qquad P\Leftrightarrow m$

$~U_{p_i}U_{p_j}U_{p_k}\Leftrightarrow \sigma _i\sigma _j\sigma _{k\qquad
}\qquad \quad \qquad ~\qquad \quad \qquad \beta _1,\beta _3\Leftrightarrow
h,J_3\qquad $
\end{center}
This duality becomes obvious if one construct the dual lattice to the 
generalized Bethe lattice of plaquettes by joining nearest centers of 
plaquettes to each other. Indeed, as a result we get the Husimi tree 
(Fig. 2).  

The plot of the $P$ for different values of $\beta _1$, $\beta _3$ 
( $\beta _3 = -1/{kT} $) is presented in Fig 3 in the thermodynamic limit. 
For high $\beta _3$ one has a stable fixed point for $P$. 
Decreasing  $\beta _3$ one can receive a periodic orbit of period 2 for $P$. 
Further decreasing $\beta _3$ one can get a periodic orbit of period $2^n$ or 
a chaotic attractor which does not have a stable periodic orbits. 
The last one (strange attractor) we shall call the chaotic state of 
the $Z(2)$
gauge theory which has a rich phase structure. Note that at high and low 
values of $\beta _1$, $P$ completely determines the state of the system for a 
stable fixed point but for some intermediate values of $\beta _1$ 
(corresponded to the chaotic region) we can't consider $P$ as an order 
parameter. The reason of such phase transitions is the different geometrical 
and dynamical properties of the attracting sets\cite{A24} of the map 
Eq.(\ref{R7}) at different values of $\beta _1$ and $\beta _3$. The relevant 
information about the geometrical and dynamical properties of attractors can 
be found from the generalized dimensions \cite{R12} or Lyapunov exponent. To 
be a good order parameter, a quantity must be capable of describing and 
characterizing qualitative changes at the bifurcation points \cite{RR14}. 
Thus for characterization of the $Z(2)$ gauge model on the generalized Bethe 
lattice in the chaotic region we have to consider the generalized dimensions
or Lyapunov exponents as an order parameter. Recently, we have calculated the
Lyapunov exponent of map Eq.(\ref{R7}) in case of the fully developed chaos 
and shown its nonanalitic behavior\cite{ADKO}.

Note that the phase structure of $Z(2)$ model on generalized Bethe lattice
allow one to take continuum limit at rich sets of $\beta _1$, $\beta _3$ of 
second order phase transitions. In particular, we solve numerically the 
following system of equations 
$$  
\cases{f(x)-x=0 \cr f^{'}(x)=-1\cr}, 
$$
which determines the points of the first period doubling bifurcation, i.e. 
the points of the second order phase transition from the disordered to the 
two-sublattice ordered phase. The result is presented in Fig.4. 

In summary we have shown that $Z(2)$ model on generalized Bethe lattice
exhibits cascade of phase transition according to Feigenbaum scheme in the
presence of three plaquette interaction. The duality between this gauge model 
and three site Ising spin model on Husimi tree is shown. We obtained chaotic 
states at some values of parameter at which order parameter exhibits chaotic 
behavior. These initial results have opened new challenges for theories of 
stochastic processes, especially in the direction of stochasticity of vacuum 
in QCD. We hope that the investigation of the $Z(3)$ gauge theory in 
the same approach will give a possibility to find out information about
confinement-deconfinement phase transition and quark-gluon plasma formation
\cite{AA7}.

The authors are grateful to R. Flume, S. Ruffo, and A. Sedrakian for useful
discussions.

This work was partly supported by the Grant-211-5291 YPI of the German
Bundesministerium fur Forshung and Technologie and by INTAS-93-633.

\newpage

\begin{center}
Figure Captions
\end{center}

\vspace*{2cm}

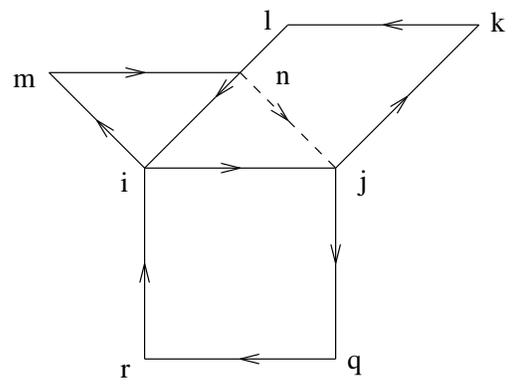
Fig.1. a - Generalized Bethe lattice of plaquettes. b - The minimal product 
of the gauge variable along the three plaquettes. The arrows show the bypass 
routing of loop.

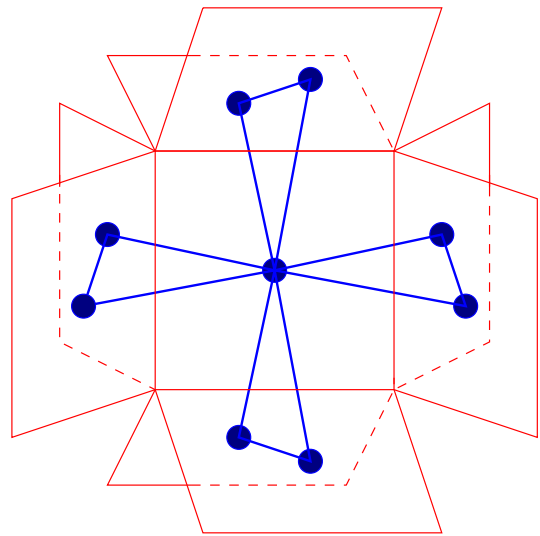
Fig.2. The duality between the generalized Bethe lattice of plaquettes and 
the three site Husimi tree. The sites are denoted by circles. 

Fig.3. Plots of $P$ versus $\beta_1$ for different temperatures $T$ 
($\beta_3 = -1/{kT}$). a - $T= 2$, b - $T= 1$, c - $T=0.6$, d - $T=0.3$.

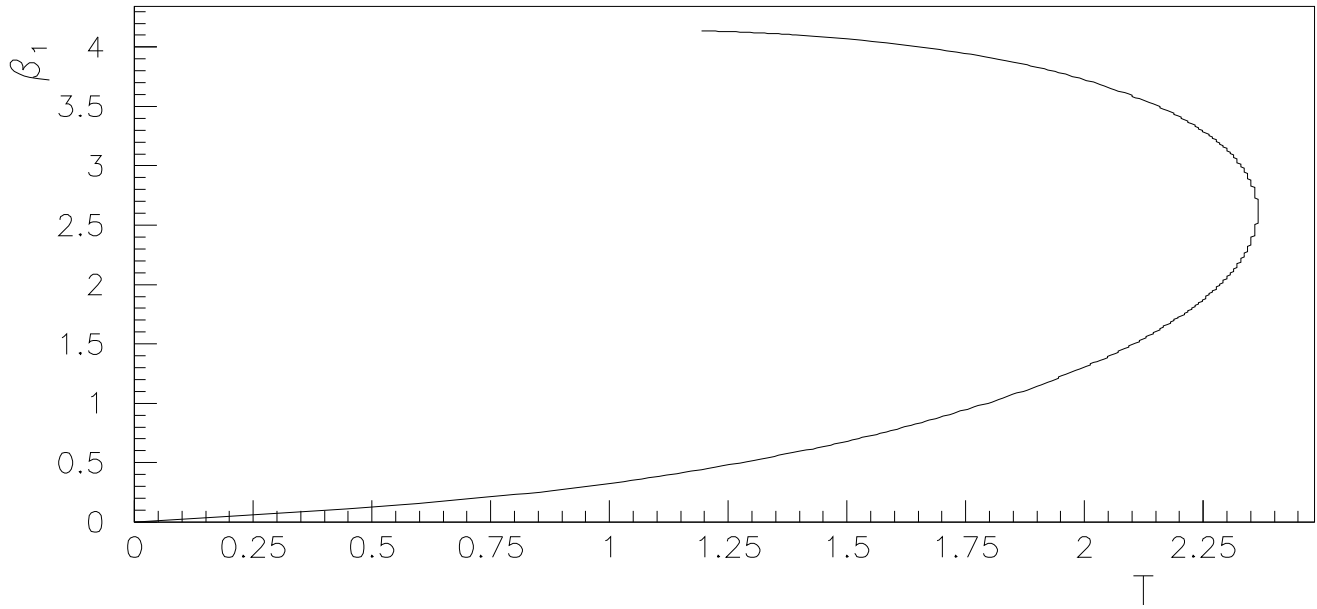
Fig.4. The line of the second order phase transitions from the disordered 
to the two-sublattice ordered phase corresponded to the first points of the 
period doubling bifurcation.

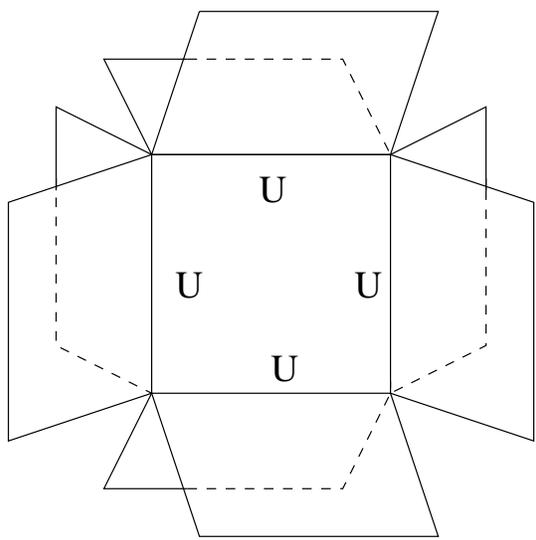
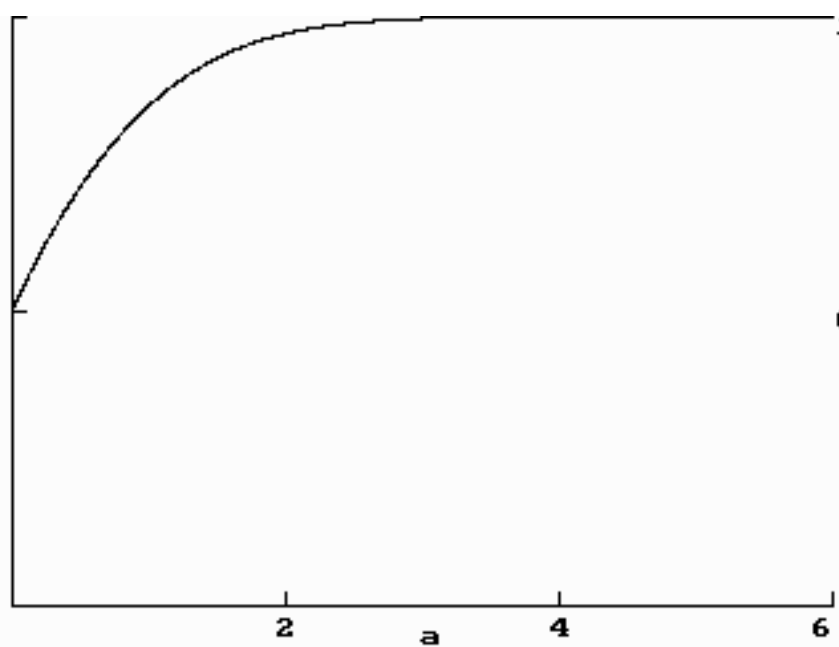
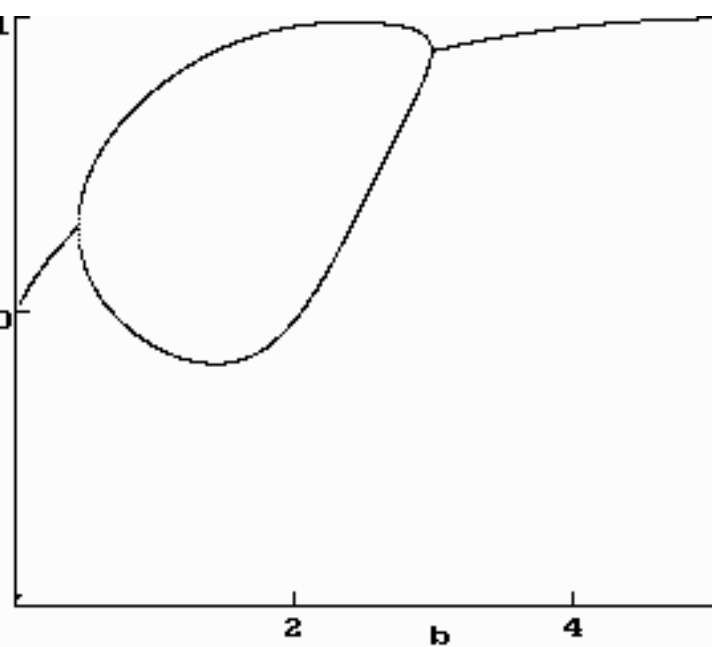
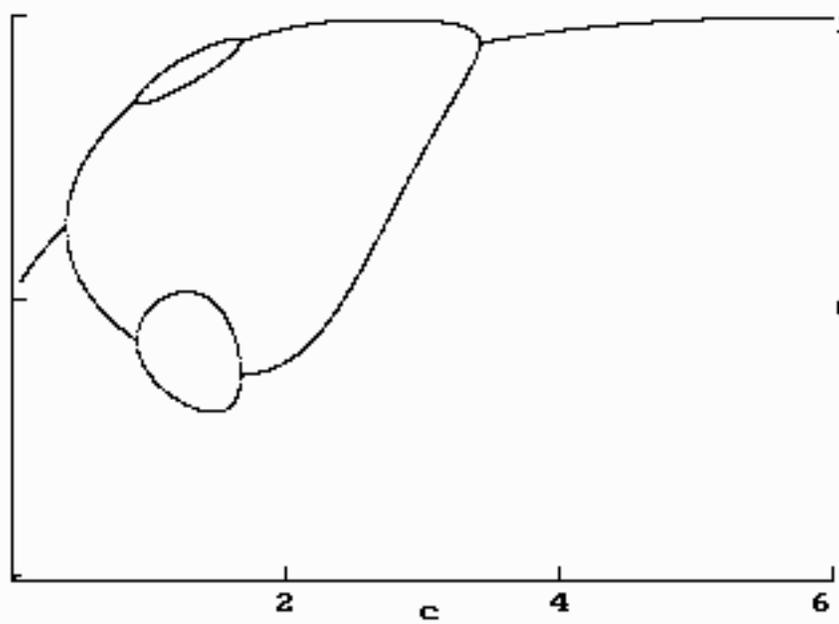
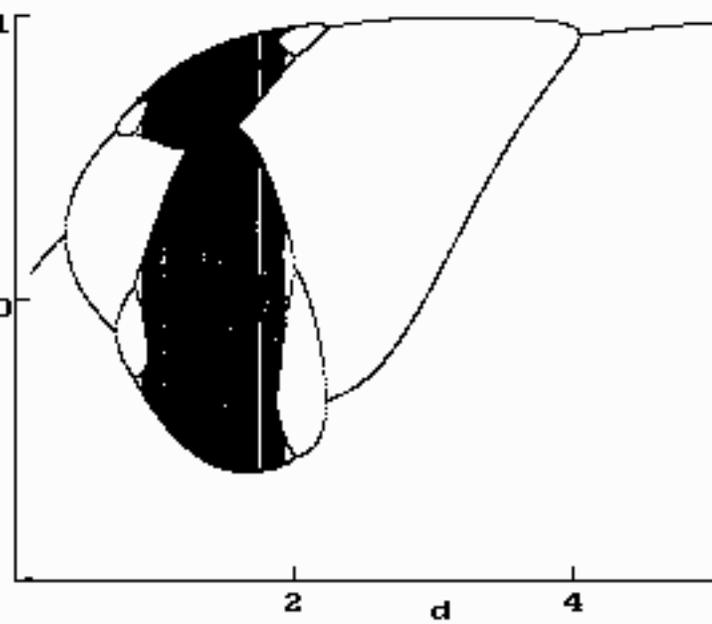
\end{document}